\documentclass[aps,prl,10pt,reprint,superscriptaddress,showpacs]{revtex4-1} 

\usepackage{amsmath}     
\usepackage{amssymb}     
\usepackage{graphicx}    
\usepackage{hyperref}    
\usepackage{color}

\begin{document}

\title{Proposed Detection of the Topological Phase in Ring-Shaped Semiconductor-Superconductor Nanowires Using Coulomb Blockade Transport}

\begin{abstract}
In semiconductor-superconductor hybrid structures a topological phase transition is expected as a function of the chemical potential or magnetic field strength. We show that signatures of this transition can be observed in nonlinear Coulomb blockade transport through a ring shaped structure. In particular, on the scale of the superconducting gap and for a fixed electron parity of the ring, the excitation spectrum is independent of flux in the topologically trivial phase but acquires a  characteristic $h/e$ periodicity in the  nontrivial phase. We relate the $h/e$ periodicity to the recently predicted $4\pi$ periodicity of the Josephson current across a junction formed by two topological superconductors. 
\end{abstract}

\author{Bj\"orn Zocher}
\affiliation{Institut f\"ur Theoretische Physik, Universit\"at Leipzig, D-04009 Leipzig, Germany}
\affiliation{Max-Planck-Institute for Mathematics in the Sciences, D-04103 Leipzig, Germany}

\author{Mats Horsdal}
\affiliation{Institut f\"ur Theoretische Physik, Universit\"at Leipzig, D-04009 Leipzig, Germany}
\affiliation{Max-Planck-Institute for Solid State Research, Heisenbergstra{\ss}e 1, D-70569 Stuttgart, Germany}

\author{Bernd Rosenow}
\affiliation{Institut f\"ur Theoretische Physik, Universit\"at Leipzig, D-04009 Leipzig, Germany}

\date{September 10, 2012}


\maketitle

\textit{Introduction}.---The investigation of topological phases of quantum systems has become one of the most exciting developments in the condensed matter community. Of particular interest are the topological properties of wave functions (WFs) and exotic quasiparticles~\cite{HK2010,QZ2010}. For this reason, much effort has been invested in the study of topological superconductors (TSCs), which have been predicted to host Majorana fermions~\cite{RG2000,FK2009,SF2009,V2009,SLTS2010,LTYSN2010,CF2011,STLSS2011,WSBT2011}. One of the defining properties of a topologically ordered state is the ground state degeneracy  on surfaces with nonzero genus. In particular,  the grand canonical ground state of the $p_x+ip_y$ (nontrivial) TSC on the torus strongly depends on  boundary conditions (BCs) for each of the two fundamental cycles~\cite{RG2000,OKSNT2007}. The three ground states with at least one antiperiodic BC are all described by even parity WFs, while the ground state with only periodic BCs shows an odd parity ground-state WF. In contrast, the ordinary $s$-wave (trivial) SC on the torus possess a fourfold degenerate ground state with an even parity~\cite{OKSNT2007}. 

In this Letter, we consider a ring shaped one-dimensional SC in the limit where the gap $\Delta$  is much larger than the single-particle level spacing $d$. In the Coulomb blockade regime with a fixed particle number $N$, the degeneracy of grand-canonical ground states on the torus is reflected in the excitation spectrum, which can be observed in nonlinear transport \cite{BRT1996,DR2001}. In a trivial SC, the lowest excitation above a ground state with even $N$  breaks a Cooper pair and hence costs the energy $\delta E \approx 2 \Delta$. When changing BCs by varying the flux through the ring, $\delta E$ oscillates  with a small amplitude  $d^2/\Delta$, i.e.~is essentially flux independent \cite{fluctuation}. The ground state for odd $N$ has an unpaired particle, and hence $\delta E \approx d^2/\Delta$ with oscillations of the same magnitude. For nontrivial TSCs however, ground states without an unpaired particle have even $N$ for anti-periodic BCs, and odd $N$ for periodic BCs. As a consequence, $\delta E$ oscillates between $d^2/\Delta$ and $2 \Delta$ with a flux period of $h/e$, very different from the trivial case. As these conclusions only rely on the existence of a superconducting gap $\Delta > d$, they should be robust against disorder \cite{LSS2011,Brouwer+11}.

One promising candidate for TSCs are semiconductor (SM) nanowires with strong Rashba spin-orbit coupling in a magnetic field and proximity coupled to an $s$-wave SC~\cite{LSS2010,ORO2010,A2010,AOROF2011}. Detection schemes for the observation of Majorana fermions in TSCs using the periodicity of the Josephson effect~\cite{K2001,FK2009,LPAROO2011,HHAB2011}, tunneling spectroscopy~\cite{LLN2009,F2010,WADB2011,LF2011,LB2011}, interferometry~\cite{ANB2009,GS2011}, and transport signatures~\cite{ADHWB2011,TSSZS2011,LT2011} have been suggested. The robustness of the $h/e$-periodic Josephson effect against a Coulomb charging energy larger than the Josephson energy was demonstrated in Ref.~\cite{HHAB2011}. Here, we go significantly beyond these results. We suggest using a large Coulomb charging energy as a tool to force the hybrid system into a state with fixed parity. In this regime, we use an unbiased numerical minimization to  calculate the excitation energies as a function of flux and particle number parity. The spectra show clear signatures of both the trivial SC and nontrivial TSC phase as expected from the general discussion above. The transition between the trivial and nontrivial phase  gives rise to the closing and reopening of an excitation gap. Finally, we compare the flux periodicity of the excitation spectra with the $4\pi$ periodicity of a Majorana ring with one weak link~\cite{K2001}. 

\begin{figure}[t]
\includegraphics[width=0.45\textwidth]{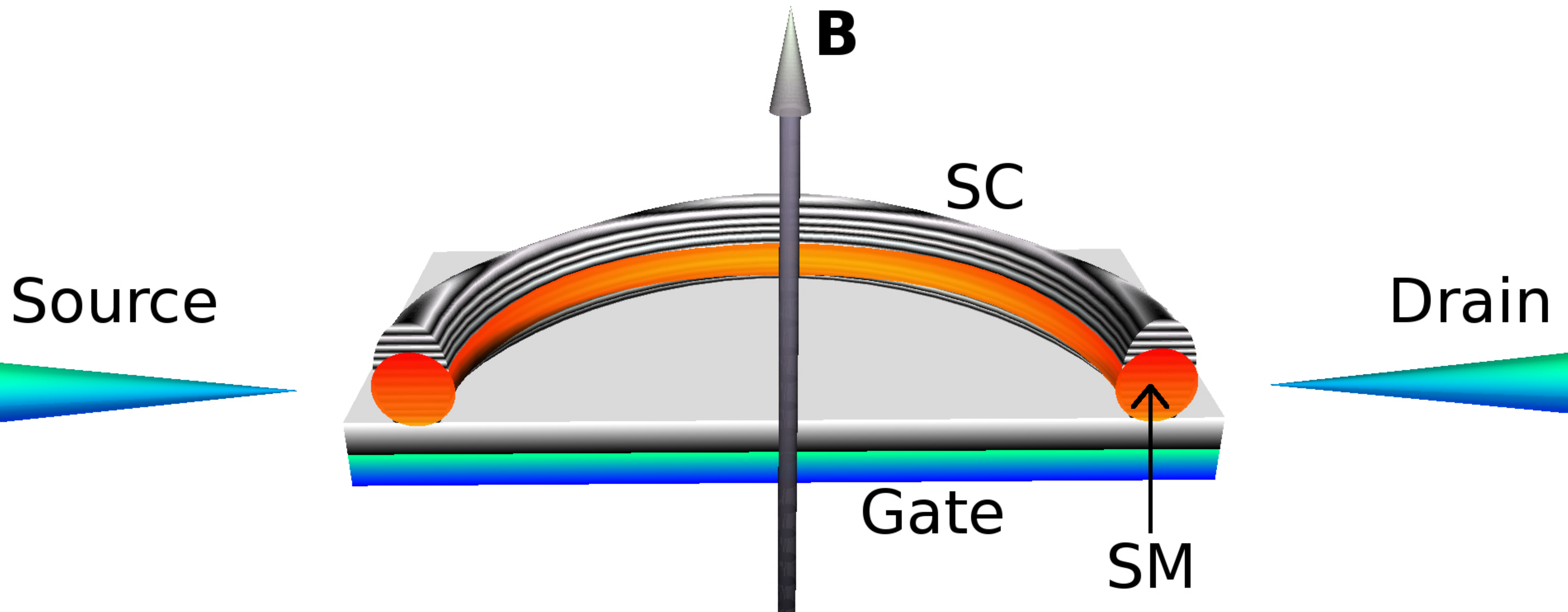}
\caption{(Color online) Cross section of the experimental setup for a ring shaped SM/SC hybrid system. The SC is sputtered on top of the SM which itself is deposited on a gate electrode. } 
\label{fig:setup}
\end{figure}

\textit{Model system}.---We consider a SM nanowire with strong spin-orbit coupling forming a loop of radius $R$, separated from a gate electrode by a thin insulating layer. On top of the nanowire a proximity coupled  $s$-wave SC is deposited, see Fig.~\ref{fig:setup}. Tunneling into and out of the SM/SC hybrid system is possible via source and drain electrodes. Assuming a strong capacitive coupling between the nanowire and SC, the Coulomb energy of the hybrid system is given by
%
\begin{equation}
H_C=E_C(N+N_{SC})^2 - eV_G(N+N_{SC}),
\label{eqn:Charge}
\end{equation}
%
where $E_C$ denotes the charging energy, $V_G$ the gate potential, and $N$ ($N_{SC}$) the number of excess electrons in the SM (SC) attracted by the gate voltage. The Hamiltonian Eq.~\eqref{eqn:Charge} describes the Coulomb blockade physics of the hybrid system: When the charging energy is degenerate with respect to changing $N + N_{SC}$ by one, a peak in the linear conductance through the hybrid system is observed. For nonzero source-drain voltage $V$, resonances in differential conductivity  appear when $e V/2 = E (N \pm 1) - E^\mathrm{gs} (N)$ where $E(N)$, is the total energy of an $N$-electron state and $E^\mathrm{gs}(N)$ the respective ground-state energy. The distances between these peaks are independent of the charging energy and directly give the fixed particle number excitation spectrum, $E(N) - E^\mathrm{gs} (N)$.  We assume that the excitation gap in the SC is much larger than the effective gap $\Delta_{\mathrm{eff}}$ in the SM. Then, all electrons in the SC are paired and unpaired electrons can only show up in the SM. In this regime, breaking of Cooper pairs occurs in the SM only and can be observed as resonances in the nonlinear Coulomb blockade conductance, similar to the experiment on metallic nanograins \cite{BRT1996}. Due to the charge $2e$ of Cooper pairs, Andreev tunneling is not resonant for $eV/2 < E_c-\Delta_{\mathrm{eff}}$ and can be neglected~\cite{cotunneling,HG1993,AN1990}. 

The  Hamiltonian describing the lowest energy subband of the nanowire is given by~\cite{MMK2002} 
%
\begin{align}
H=&\sum_{k\in \mathbb{Z}} \Big\{ \psi_{k\sigma}^\dagger \Big[\frac{\hbar^2}{2m^*R^2}\Big(k+\frac{\Phi}{\Phi_0} \Big)^2-\mu+\sigma \frac{g\mu_B B}{2} \Big] \psi_{k\sigma} \nonumber\\
&+\frac{\alpha}{R}  \Big(k+\frac{1}{2}+\frac{\Phi}{\Phi_0} \Big)\Big( \psi_{k\uparrow}^\dagger \psi_{k+1\downarrow}+ \psi_{k+1\downarrow}^\dagger \psi_{k\uparrow}\Big)\Big\},
\label{eqn:Hamiltonian}
\end{align}
%
where the operator $\psi_{k \sigma}^\dagger$ ($\psi_{k \sigma}$) creates (annihilates) an electron with spin $\sigma$ and angular momentum $\hbar k$, $m^*$ is the effective band mass, and $\mu$  the chemical potential. We expect the following discussion to hold also for the more general case of an odd number of occupied transverse modes~\cite{LSS2011}. The Rashba spin-orbit coupling, $\alpha$, couples states $\{|k\uparrow\rangle, |k+1\downarrow\rangle\}$ and creates two helical bands with the spin rotating within the $x$-$y$ plane. The bands cross each other at $k= -1/2 - \Phi/\Phi_0$.  The magnetic field, $B$, tilts the spin direction  out of the $x$-$y$ plane, removes the level crossing,  and opens a spin gap $E_Z=g\mu_B B/2$. $\Phi/\Phi_0$ denotes the magnetic flux through the loop in units of the flux quantum $\Phi_0=h/e$. We find the single-particle dispersion of the tilted helical bands,
%
\begin{equation}
\epsilon_{\pm,\tilde{k}}=\frac{\hbar^2\big(\tilde{k}^2+\frac{1}{4} \big)}{2m^*R^2}  \pm \sqrt{\Big(\frac{\hbar^2 \tilde{k}}{2m^*R^2}-E_Z\Big)^2 +\frac{\alpha^2 \tilde{k}^2}{R^2}},
\label{eqn:eigenenergies}
\end{equation}
%
where $\tilde{k}=k+\Phi/\Phi_0+1/2$.

The $s$-wave SC is described within the Ginzburg-Landau formalism by the free energy density 
%
\begin{equation}
f_\mathrm{GL}[|\Delta_s|,q]=f_0(|\Delta_s|^2)+\frac{\hbar^2|\Delta_s|^2}{2m_sR^2}\Big(q+\frac{2\Phi}{\Phi_0}\Big)^2 +\frac{B^2}{2\mu_0},
\label{eqn:fGL}
\end{equation}
%
where $f_0$ is the free energy for zero flux, $\Delta_s$ the pairing potential, $\hbar q$ the condensate angular momentum, and $m_s$ the mass of the Cooper pairs. Minimization of $f_{\rm GL}$ demands that $q$ is the integer nearest to $-2\Phi/\Phi_0$ and that $\delta f_\mathrm{GL} /\delta \Delta_s=0$. In the following, we neglect the small oscillations in $|\Delta_s|$ and focus on the large effect of parity and flux on the addition spectrum of the SM ring. The proximity coupling between the $s$-wave SC and the nanowire gives rise to a pairing term~\cite{STLSS2011}
%
\begin{equation}
H_{SC}= \sum_{k\in \mathbb{Z}} \Big[\Delta(\Phi) \psi_{k \uparrow}^\dagger  \psi_{-k+q \downarrow}^\dagger  +\Delta^*(\Phi)  \psi_{-k+q \downarrow}  \psi_{k \uparrow} \Big],
\label{eqn:HSC}
\end{equation}
%
which couples states $|k \uparrow\rangle$ and $|-k+q\downarrow\rangle$. As a consequence, the Hamiltonian is block diagonal, and within each block a quadruplet $\{|k\uparrow\rangle, |k+1\downarrow\rangle, |-k+q\downarrow\rangle, |-k-1+q\uparrow\rangle\}$ is coupled. For odd $q$, the quadruplet for $k=(q-1)/2$ reduces to the doublet $\{|(q-1)/2\uparrow\rangle, |(q+1)/2\downarrow\rangle \}$. The pairing potential $\Delta$, which is reduced in magnitude as compared to $\Delta_s$, plays a crucial role since it sets two excitation energies. It both opens an effective pairing gap at the Fermi surface and it modifies the Zeeman gap at $\tilde{k}=0$. For $\Delta^2>E_Z^2 -\mu^2 $ both helicities are occupied in the ground state and $\Delta$ pairs generalized time-reversed pairs at both sets of Fermi points. Hence, the nanowire is in a trivial state with SC gaps at both $\pm \tilde{k}_F$ and $\tilde{k}=0$. For $\Delta^2<E_Z^2 -\mu^2 $ on the other hand, the band structure is different in an important way because now there is a spin gap at $\tilde{k}=0$ and an SC gap only at $\pm \tilde{k}_F$~\cite{ORO2010}. If $E_Z \gg \Delta,\mu$, it is justified to only consider the lower band and to project the proximity induced singlet pairing onto that band~\cite{LSS2010,AOROF2011}. In this limit, the low-energy theory of the ring model with flux $\Phi$ can be mapped onto Kitaev's model~\cite{K2001} with periodic BC and flux $\Phi+\Phi_0/2$. The projected model contains doublets $\{|p\rangle, |-p\rangle \}$ for $\Phi/\Phi_0 \in [n-1/4,n+1/4]$ with integer $n$ and effective momentum $p=k-q/2+1/2$, whereas for $\Phi/\Phi_0 \in [n+1/4,n+3/4]$,  the doublet for $p=0$ reduces to the singlet $|p=0\rangle$.

In analogy to the generalized variational approach in Ref.~\cite{DR2001}, we consider variational WFs for the projected Hamiltonian. For each doublet, states with even and odd parity are generated by applying the operators
\begin{subequations}
\label{eqn:paritytotal}
\begin{eqnarray}
P_-(p)  &= s_p  c_p^\dagger +t_p c_{-p}^\dagger, \label{eqn:Pm}\\
P_+(p)&= u_p +v_p c_p^\dagger c_{-p}^\dagger \label{eqn:Pp}
\end{eqnarray}
\end{subequations}
%
to the vacuum state. Here the $c$ operators denote electrons of the lower helical band $\epsilon_-(p)$. General ansatz WFs for  even (odd) parity are
%
\begin{equation}
|\Psi_{\mathrm{e(o)}}\{\tau_p\}\rangle = \prod_{p\ge 0} P_{\tau_p}(p) |0 \rangle , \ \prod \tau_p =+1(-1),
\label{eqn:WF}
\end{equation}
%
where $|0 \rangle$ is the vacuum for the $c$ electrons. 
\begin{figure}    [b]
\includegraphics[width=0.45\textwidth]{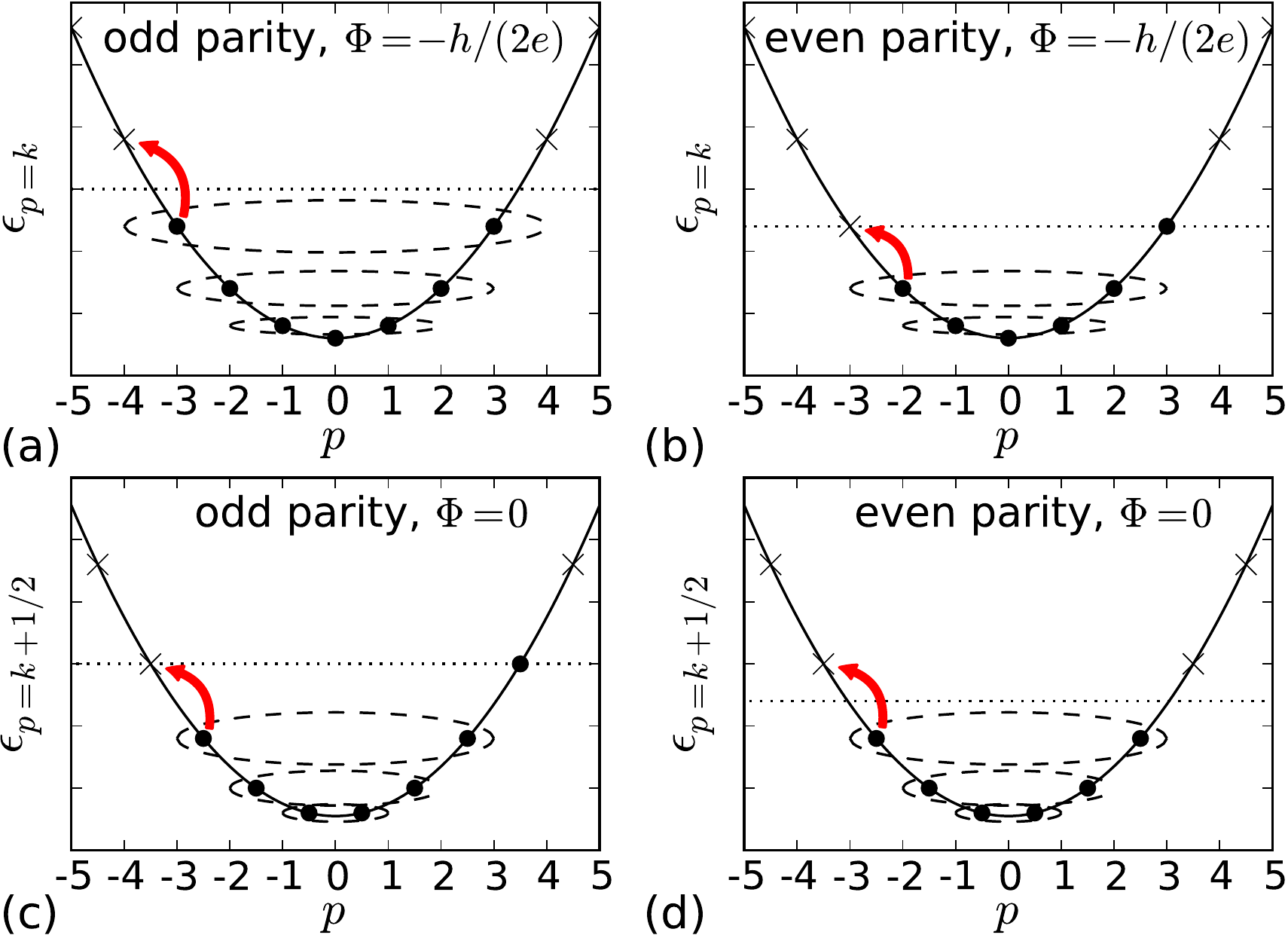}
\caption{(Color online) Sketch of the dispersion and the effective pairing for the lower helical band $\epsilon_-(p)$. The $o$ markers (x) denote the occupied (empty) single-particle levels for $\Delta=0$. The dashed ellipses illustrate the paired single-particle levels when switching on the proximity induced SC pairing potential. Arrows indicate the transport of a single quasiparticle to produce the lowest excited state. }
\label{fig:pairing}
\end{figure}
To obtain the energy spectrum for arbitrary magnetic flux, we first minimize the Ginzburg-Landau free energy Eq.~\eqref{eqn:fGL} to find the pair wave number $q$, which is then used to construct the grand canonical mean-field ansatz WFs Eq.~\eqref{eqn:WF}. For each set of $\{ \tau_p\}$, we determine the corresponding energy by unbiased minimization of $E(N,\{ \tau_p\})=\langle H \rangle +\mu_N N$ with respect to the variational parameters $(s_p, \ \dots, \ v_p)$. Here $\mu_N$ is fixed by the mean particle number $N=\langle \sum c_p^\dagger c_p\rangle$ in the SM nanowire. By rank-ordering the $E(N,\{ \tau_p\})$, we find the ground states for both even and odd parity. To obtain the excited states, we then apply the Bogoliubov operators $a_{p,1}^\dagger=u_p c_p^\dagger-v_p c_{-p} $ and $a_{p,2}^\dagger=u_p c_{-p}^\dagger+v_p c_p$ with $p > 0$ to the ground-state WF. 

In Figs.~\ref{fig:pairing}(a) and~\ref{fig:pairing}(b) we sketch a bare parabolic dispersion, the generalized time-reversed partners for $\Phi=-h/2e$, and the single-particle excitation spectrum. The ground-state WF for odd parity is given by $|\Psi_o^\mathrm{gs} \rangle = P_-(0) \prod P_+(p) |0 \rangle$, where all time-reversed partners are paired and the zero momentum electron is unpaired. The lowest excited state has two unpaired electrons at $p_F$ and $p_F+1$, $|\Psi_o^{ij} \rangle = a_{p_F,i}^\dagger a_{p_F+1,j}^\dagger |\Psi_o^\mathrm{gs} \rangle$, which shows up in a spectroscopic gap of $2\Delta_{\mathrm{eff}}$. On the other hand, the ground-state for  even parity is given by $|\Psi_e^\mathrm{gs} \rangle =  P_-(0)P_-(p_F) \prod P_+(p) |0 \rangle$ with two unpaired electrons. In contrast to the odd parity case, we find the lowest excited state by breaking the pair at $p_F-1$ and creating a new one at $p_F$,  $|\Psi_e^{ij} \rangle = a_{p_F,i}  a_{p_F-1,j}^\dagger |\Psi_e^\mathrm{gs} \rangle$. Therefore, the excitation energies for the even parity are determined by the level spacing. In Figs.~\ref{fig:pairing}(c) and ~\ref{fig:pairing}(d) we illustrate the pairing for $\Phi=0$. Here, we find that the behavior is reversed compared to the case $\Phi=-h/2e$; i.e. the ground state for the even parity contains only paired levels whereas the ground state for the odd parity has one unpaired electron at the Fermi surface.

\textit{Numerical results}.---We now consider the full Hilbert space again. In analogy to what we explained above, we define generalized operators $P_\pm(k)$ for each quadruplet of the unprojected Hamiltonian and construct the ansatz WFs as in Eq.~\eqref{eqn:WF}. We then minimize the energy $E(N)$, where $N=\langle \sum \psi_{k\sigma}^\dagger \psi_{k\sigma}\rangle$, to obtain the ground state~\cite{DR2001}. The lowest excited states are again given by pairwise creation of Bogoliubov quasiparticles near the Fermi surface. We note that in the coupled  SM/SC system, the number of electrons in the nanowire is not a good quantum number and the use of grand canonical WFs is fully justified. We have verified that the excitation spectrum depends smoothly on the mean particle number $N$. 

Both InAs and InSb were proposed to be suitable semiconducting materials due to a strong spin-orbit coupling~\cite{LSS2010,ORO2010}. For $R=0.5 \, \mu \mathrm{m}$, characteristic values for these  materials are $\hbar^2/(2m^*R^2) = 0.002 \,\mathrm{meV}$, level spacing at the Fermi energy $d = 0.08 \, \mathrm{meV}$,  $g\mu_B=2\, \mathrm{meV/T}$, and $\alpha /R=0.02 \, \mathrm{meV}$~\cite{WADB2011}. Furthermore, we consider a  proximity potential $\Delta=0.5 \, \mathrm{meV}$ which leads for $E_Z=1\ \mathrm{meV}$ and $\mu=0$ to an effective pairing gap of $\Delta_{\mathrm{eff}} \approx 0.2\, \mathrm{meV}$ \cite{suppression,OS2004}. To ensure single-electron tunneling through the SM/SC system, we consider the case $E_C \gg\Delta_{\mathrm{eff}}$. 
\begin{figure}[tb]
\includegraphics[width=0.48\textwidth]{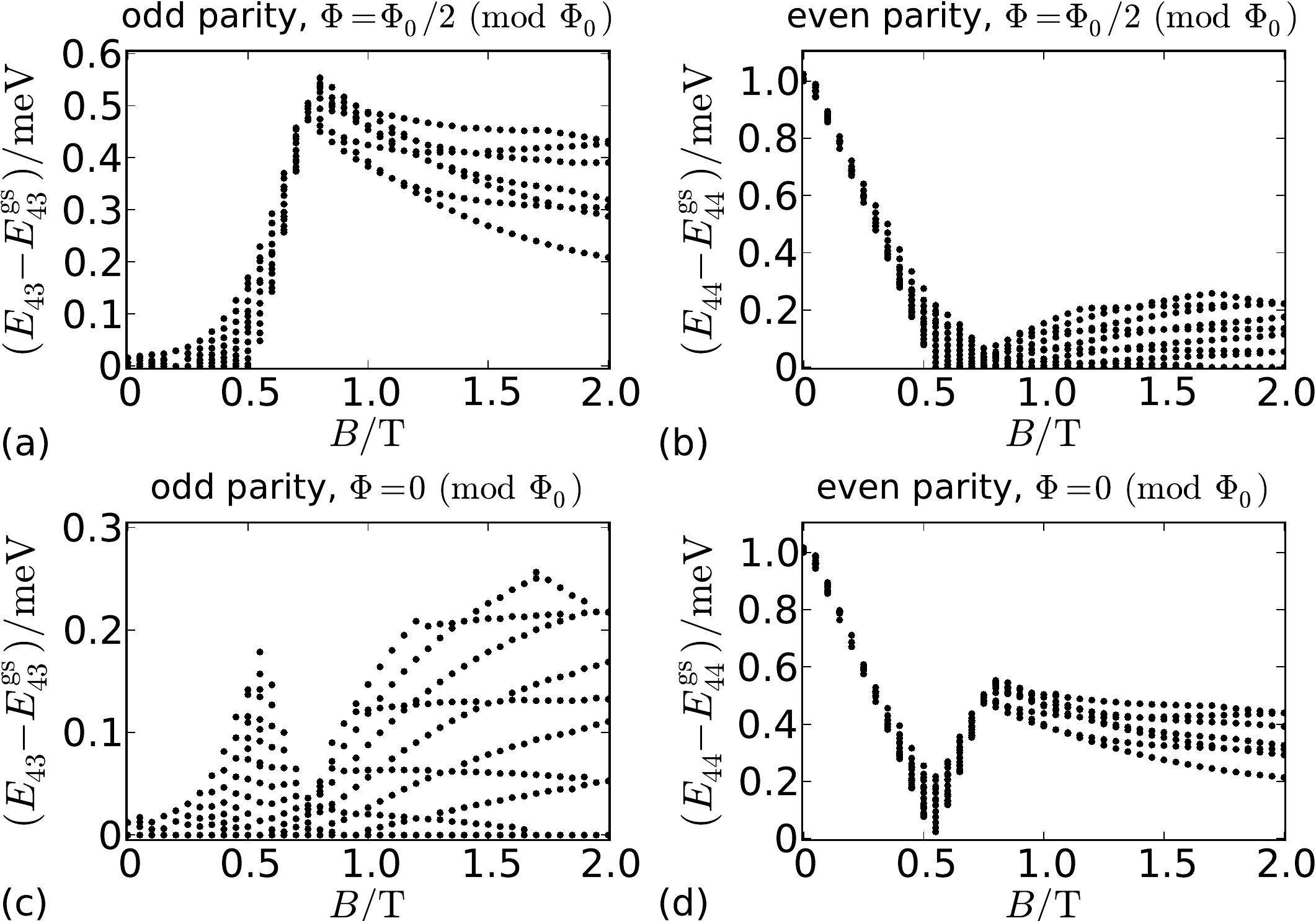}
\caption{Magnetic field dependence of the energy differences. $B$ is varied in discrete steps with the flux always being a (half-) integer multiple of $\Phi_0$. The lowest excited states in the nontrivial phase ($B>0.5$ T)  are sketched in Fig. \ref{fig:pairing} for the projected model. }
\label{fig:magnetic}
\end{figure}

The external magnetic field  $B$ drives the hybrid system  through a topological phase transition. In the following, $B$ is varied in discrete steps with the flux always being a (half-) integer multiple of $\Phi_0$, such that the only effect is a variation of the Zeeman energy. Figure~\ref{fig:magnetic} shows excitation energies as a function of $B$ for several combinations of magnetic flux and parity. We see  qualitative differences between  the trivial phase of the nanowire for $B\lesssim 0.5 \, \mathrm{T}$ and the nontrivial phase for $B\gtrsim 0.5 \, \mathrm{T}$~\cite{STLSS2011}. For $B\lesssim 0.5 \, \mathrm{T}$, results are typical  for SC in ultrasmall metallic grains~\cite{BRT1996, DR2001}: for even electron parity, the excitation spectrum displays a large spectroscopic gap $\sim 2 \Delta_{\mathrm{eff}}$, whereas no such gap appears for an odd  parity, independent of magnetic flux. The origin of the large gap for even parity is that all excitations break a Cooper pair, while for odd parity the ground state has one unpaired electron and therefore the lowest excitation energies are determined by the level spacing as $d^2/\Delta_{\mathrm{eff}}$ \cite{fluctuation}. 

For $B\gtrsim 0.5 \, \mathrm{T}$ we observe a strikingly different parity effect, and find  that the excitation energies depend on both magnetic flux and electron parity. In Figs.~\ref{fig:magnetic}(a) and (d) we find a spectroscopic gap that originates from breaking a pair [compare to illustrations Figs.~\ref{fig:pairing} (a) and (d)], which costs the energy $2\Delta_{\mathrm{eff}}$. In contrast, the excitation energies in Figs.~\ref{fig:magnetic}(b) and~\ref{fig:magnetic}(c) are determined by the level spacing, compare to illustrations Figs.~\ref{fig:pairing} (b) and (c). The topological phase transition is mirrored by the closing and reopening of the excitation gap; see Fig.~\ref{fig:magnetic}(d).

\begin{figure}[tb]
\includegraphics[width=0.45\textwidth]{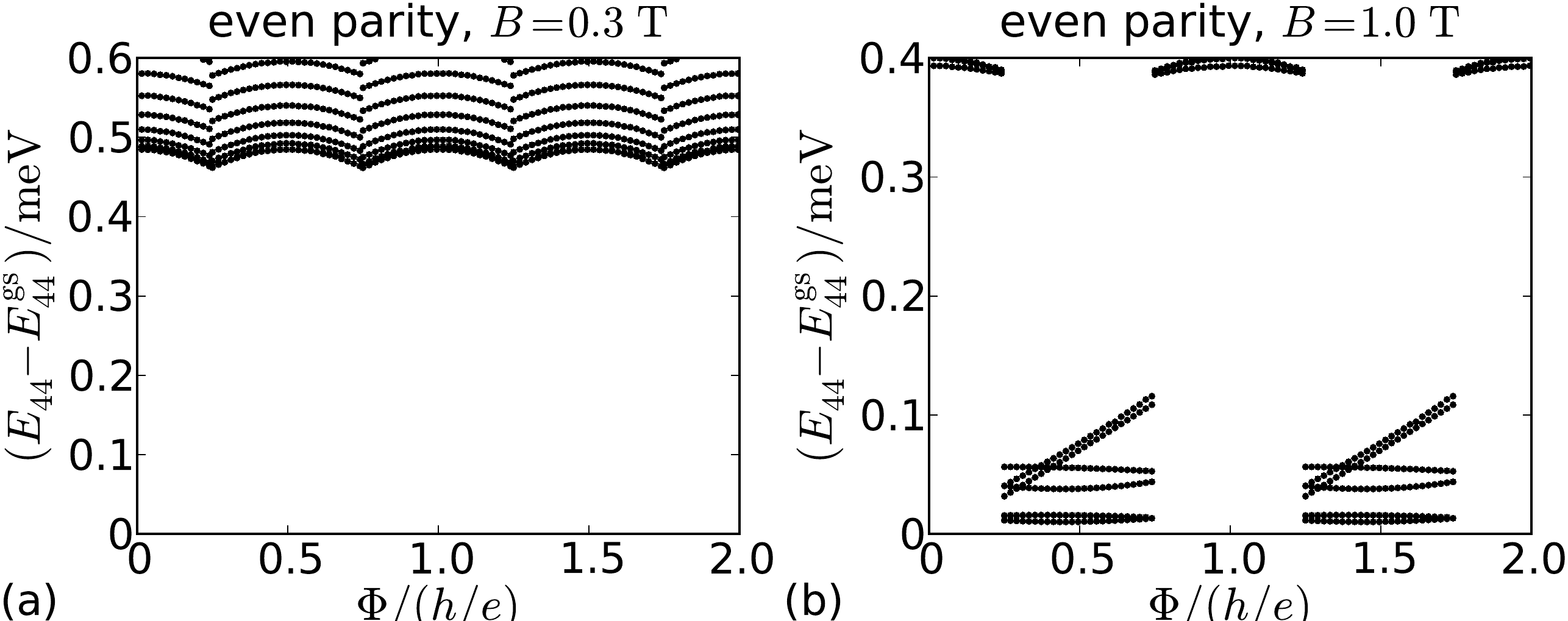}
\caption{Energy differences as function of the magnetic flux, (a) for $B=0.3\, \mathrm{T}$,  and (b)  for $B=1.0\, \mathrm{T}$. Not all higher energies are shown.}
\label{fig:flux}
\end{figure}

In Fig.~\ref{fig:flux}, excitation energies  as a function of  magnetic flux for both  trivial ($B=0.3\,\mathrm{T}$) and nontrivial sectors ($B=1.0\,\mathrm{T}$) are shown for even parity. In the trivial phase, they are of order $2 \Delta_{\mathrm{eff}}$ with small  $\Phi_0/2$ periodic oscillations of order $d^2/\Delta_{\mathrm{eff}}$; see Fig.~\ref{fig:flux}(a).  For the odd parity case (not shown), they  are determined by the level spacing. In the nontrivial phase however, large oscillations with period $\Phi_0$ and amplitude $2 \Delta_{\mathrm{eff}}$ are found; see Fig.~\ref{fig:flux}(b): The excitation energies for $\Phi/\Phi_0 \in (1/4,3/4)$ are determined by the level spacing, while they display the effective gap $2\Delta_{\mathrm{eff}}$ for $\Phi/\Phi_0 \in (3/4,5/4)$ due to the pairwise creation of Bogoliubov quasiparticles. For odd parity,  we qualitatively find the same spectrum but shifted by $\Phi_0/2$, as follows from the earlier discussion. All these results back up the general arguments in the introduction, connecting ground state degeneracies on the torus to parity and flux periodicities of excitations. 

We now  relate the $\Phi_0$ flux periodicity  in the nontrivial phase to the recently discovered $4 \pi$ periodicity of the Josephson current between two TSCs~\cite{K2001,FK2009,HHAB2011}. To leading order in the tunnel coupling, the Josephson energy between two 1D TSCs is given by
%
$H_J(\Delta\phi)=i \gamma_1\gamma_2  \Gamma \cos\Big(\frac{\Delta \phi}{2}\Big)$,
%
where $\gamma_1$, $\gamma_2$ are  operators for the end Majorana states connected by the junction, $\Gamma$ is the tunneling amplitude, and $\Delta \phi$ the phase difference between the SCs. The operator $i \gamma_1 \gamma_2$ with eigenvalues $\pm 1$ describes the parity of the neutral fermion state  shared between the two Majoranas.  For a fixed parity, $H_J$ has a period of $4 \pi$. When inserting the Josephson junction into a ring structure, the phase difference between the two ends is related to a flux through the ring via $\Delta \phi = \Phi/\Phi_0$, and the $4 \pi $ phase periodicity is equivalent to a $\Phi_0$ flux periodicity. If the parity is not fixed, a change of $\Delta \phi$ by $2\pi \sim \Phi_0/2$ will change the occupancy  $(i\gamma_1\gamma_2 +1)/2$ of the neutral fermion and hence the ground state parity. This is in full analogy with our finding that in the nontrivial phase the parity of the ground state changes (if coupled to a reservoir) when changing the flux through the ring by $\Phi_0/2$. Since occupying the neutral fermion describes a change in the parity of the pairing WF and not in the mean number of (charged) particles, the term ``neutral fermion'' is appropriate. 

\textit{Conclusion}.---We have investigated the signatures of Coulomb blockade transport through a SM/SC hybrid nanoring, and have shown that peculiar parity and flux periodicity effects in the excitation spectrum mirror the distinct ground state degeneracies of trivial and nontrivial SCs on the torus. The excitation spectrum provides a clear signature of the topological phase transition, and the $h/e$ flux periodicity of  excitation energies in the nontrivial phase is reflected in the $4\pi$ periodicity of the Josephson Hamiltonian for a tunnel junction between two 1D $p+ip$ TSCs. 

We acknowledge helpful discussion with T.~Hyart and A.P.~Schnyder, and financial support by BMBF. 



\begin{thebibliography}{99}
\bibitem{cotunneling}In addition, we find that Cooper pair cotunneling can be neglected if the dimensionless conductance $G$ for tunneling into the normal state SC satisfies $G \lesssim \sqrt{N_\perp}/3$, where $N_\perp$ is the number of  transverse channels in the SC \cite{HG1993,AN1990}. To satisfy this condition, we propose to not cover the nanowire with superconducting material in the vicinity of the electrodes.
\end{thebibliography}
\end{document}